\documentclass[pre,twocolumn,showpacs]{revtex4}
\usepackage{epsfig}
\begin{document}
% LyX 1.3 created this file.  For more info, see http://www.lyx.org/.
% Do not edit unless you really know what you are doing.
%\documentclass[twocolumn,english,pre,showpacs]{revtex4}
%\title{Exact solution for the influence of spectral diffusion on
%single molecule photon statistics
\title{Super and sub-Poissonian
photon statistics for
single molecule spectroscopy
}
\author{Yong He$^{1}$, Eli Barkai$^{1,2}$}
\affiliation{\ $^{1}$ Department of Chemistry and Biochemistry, Notre Dame University, Notre Dame, IN 46556. \\
\ $^{2}$ Department of Physics, Bar Ilan University, Ramat Gan 52900, Israel}
\date{\today}
\begin{abstract}
We investigate the distribution of the number of photons emitted
by a single molecule undergoing a spectral diffusion process
and interacting with a continuous wave laser field.
The spectral diffusion is modeled based on a stochastic
approach, in the spirit of the Anderson-Kubo line shape theory.
Using a generating function formalism we solve
the generalized optical Bloch equations, and obtain
an exact analytical formula for the line shape and
Mandel's Q parameter. 
The line shape exhibits well known behaviors, including motional
narrowing when the stochastic modulation is fast, and 
power broadening. 
The Mandel parameter, describing the line shape fluctuations, 
exhibits a transition from a Quantum sub-Poissonian behavior
in the fast modulation limit, to  
a classical super-Poissonian behavior found in the slow modulation
limit. 
Our result is applicable for  weak and strong laser field,
namely for arbitrary Rabi frequency. 
We show how to choose the Rabi frequency
in such a way that the Quantum sub-Poissonian
nature of the emission process becomes strongest.
A lower bound on $Q$ is found, and simple limiting behaviors
are investigated. A non-trivial behavior is obtained in the intermediate
modulation limit, when the time scales for spectral diffusion
and the life time of the excited state, become similar. 
A comparison is made between our results,
and previous ones derived based on the semi-classical
generalized Wiener--Khintchine
formula.

\end{abstract}


\maketitle

\section{Introduction}
                                                                                                                                                             
Physical, Chemical, and
Biological systems are investigated in many laboratories
using single molecule
spectroscopy  \cite{MO}.
The investigation
of the distribution of the number of photons emitted
from a single molecule source is the topic of extensive theoretical research
e.g. 
\cite{Plakh,Schenter,Berez,Molski,Haw,Jung1,Jang,WangM,Oijen,Grigolini,Barsegov2,Cao,Gennady}
and \cite{BarkaiRev} for a review. 
Since optical properties of  single molecules are 
usually very sensitive to dynamics and statics of their
environment, and since the technique removes
the many particle  averaging found in conventional measurement
techniques, single molecule spectroscopy reveals interesting
fluctuation
phenomena. 
An important mechanism responsible for the fluctuations
in the number of photons emitted from a single molecule
source is spectral diffusion e.g. 
\cite{Ambrose,Bas,Moer,Boiron,Shim2}.
In many cases
the absorption frequency
of the molecule will randomly change due to different
types of interactions between the molecule and
its environment 
(e.g. \cite{BarkaiRev,Reilly2,Geva,Bordat1,Tanimura,PRL,ADV} and Ref. therein).
For example for single molecules embedded
in low temperature glasses, flipping two
level systems embedded in the glassy environment,
induce stochastic spectral jumps in the absorption
frequency of the single molecule under investigation 
\cite{Boiron,Geva,Kador}.
In this way the molecule may come in and out of resonance with
the continuous wave laser field with which it is interacting.

Obviously a second mechanism responsible for fluctuations
of photon counts is the quantum behavior of the
spontaneous emission process \cite{Wolf,Plenio}.
In his fundamental work Mandel \cite{Mandel1} showed that a single
atom in the process of resonance fluorescence, {\em  in the absence of
spectral diffusion},
exhibits sub-Poissonian photon statistics \cite{Short}.
Photon statistics is characterized by
Mandel's $Q$ parameter
\begin{equation}
Q= {  \overline{N^2}  - \overline{N }^2 \over \overline{N}} -1
\end{equation}
where $N$ is the number of emitted photons within a certain time
interval. The case $Q<0$ is called sub-Poissonian behavior,
while $Q>0$ is called super-Poissonian. Sub-Poissonian statistics
has no classical analog \cite{Mandel1}.
Briefly, the effect is
related to anti-bunching of photons emitted from a single source
and to Rabi-oscillations of the excited state population
which favors an emission process
with some periodicity in time (see details below).
Sub-Poissonian photon statistics and
photon anti-bunching were measured in several
single 
molecule, and single quantum dots experiments
\cite{Or,Zu1,Lou,Messin,Treussart,Zwiller,Huser}.
While sub-Poissonian statistics is well understood in the
context of resonance fluorescence of an isolated electronic
transition of a simple atom in the gas phase, our theoretical understanding
of sub-Poissonian
statistics for a molecule embedded in a fluctuating
condensed phase environment
is still in its infant stages. 

 In this paper we
obtain an exact analytical
expression for the $Q$ parameter in the long time limit, 
for a single molecule
undergoing a stochastic spectral diffusion process.
To obtain the exact solution
we use the Zheng-Brown generating function method 
for single molecule photon statistics
\cite{Brown,Brown1,Brown2}. 
For the spectral diffusion we use a simple stochastic approach,
in the spirit of the Kubo--Andersen line shape theory  \cite{AK,Book}.
The model exhibits generic behaviors of line shapes
of molecules embedded in a condensed phase
environment,
e.g. motional narrowing when the stochastic
fluctuations are fast,
power broadening etc. We show that  the 
$Q$ parameter exhibits rich types of
behaviors,
in particular it reveals the quantum nature of the emission
process in the sub-Poissonian regime, while the corresponding
model line-shape
exhibits a classical behavior. 
A brief summary of our results was recently published \cite{YongPRL}.

Our analytical expressions for $Q$ classify the
transitions between sub and super Poissonian statistics.
They  give the conditions
on the spectral diffusion time scale
for sub-Poissonian behavior. 
Motional narrowing type of effect is revealed also for 
the $Q$ parameter. 
Our exact result is valid for weak
and strong excitation (i.e arbitrary Rabi frequency).
It yields the lower bound on $Q$.
The solution shows how we may choose the Rabi
frequency so that the quantum nature
of the photon emission process becomes larger,
namely how to minimize $Q$ in the Sub-Poissonian
regime. This is important for the efficient
detection of quantum effects
in single molecule spectroscopy, since choosing too small or
too large values of the Rabi frequency results in very small
and hence undetectable values of $Q$.

Finally our exact result is used to test the domain
of validity of the generalized Wiener Khintchine
which yields $Q$ in terms of a Fourier transform of
a three time dipole correlation (as well known the Wiener Khintchine
theorem yields the line shape in terms of a one time dipole
correlation function).
The theorem 
\cite{PRL,ADV} is based on
the semi-classical theory of interaction of light with matter,
and on linear response theory (i.e., weak Rabi frequency),
it yields $Q>0$.  As pointed out in \cite{BarkaiRev,ADV,Brown1} 
such a behavior
is expected to be valid only for slow enough spectral diffusion
processes. 

\section{Introduction to sub-Poissonian Statistics} 
\label{SecIntroSub}

 We briefly explain some 
of the main ideas of sub-Poissonian statistics. 
The general idea is that the photons emitted from
a single particle, e.g. a  molecule,
a nano-crystal or atom are correlated in time.
Consider first a
hypothetical  molecule, interacting with an exciting
laser field, which  emits photons with a constant time
interval $\tau$ between successive emission events.
Then
$\overline{N} = t/\tau$, 
$\overline{N^2} =  
\overline{N}^2 $, and hence $Q= -1$. Due to quantum
 uncertainty the  photon emission process
is always random and therefore $-1<Q$.
Sub-Poissonian behavior where $-1<Q<0$ 
implies that 
the stream of photons emitted from a single source
maintain correlations in their arrival times to
a detector.   

 Usually  when many molecules interact with a continuous wave laser
the emission events are not correlated, and the
fluorescence
exhibits Poissonian statistics $Q =0$. In contrast a single
molecule, once it emits a photon, is collapsed to its  ground state.
Hence immediately after an emission event the molecule cannot emit
a second time (it has to be re-excited by the laser). Hence  
successive 
photons emitted from a single molecule, seem to repel each other
on the time axis, a non-Poissonian behavior.
 This well known effect is called anti-bunching \cite{Car,Kimble}
which  is related
to sub-Poissonian statistics. 

 A second effect related to sub-Poissonian
behavior are Rabi-oscillations. 
    Consider a simple atom in the process
 of resonance fluorescence. When the electronic
 transition (frequency $\omega_0$) is in resonance with
 a continuous wave laser field (frequency $\omega_L$) the electronic transition
 can be approximated by a two level system. First let us
 mentally switch off the spontaneous emission, i.e. set the inverse
 life time of the transition $\Gamma=0$. For zero detuning 
 $\omega_L = \omega_0$ the transition will exhibit well
 known Rabi oscillations: the population of the excited
 state will behave like 
 $\rho_{ee} = \sin^2 \left( \Omega t /2 \right)$.
 Since the population in the excited state attains its maximum (minimum)  
 periodically, also the emission times of successive
 photons maintain certain degree of periodicity 
 in time, which implies sub-Poissonian statistics.  Mandel
 showed \cite{Mandel1},
 that for a two level atom in the process
 of resonance fluorescence
 \begin{equation}
Q = - {6 \Omega^2 \Gamma^2 \over \left( \Gamma^2 + 2 \Omega^2 \right)^2 }.
\label{eqMold}
 \end{equation}
 When $\Omega \ll \Gamma $ we have $Q\to 0$ since then
 successive photon emission times are not correlated,
 because the time
 between successive emissions becomes very large. 
 While when $\Omega \gg \Gamma$ the excited state is
 populated swiftly, 
 and only the finite spontaneous emission rate delays
 the emission, hence
 $Q\to 0$ also in this case. 
Using Eq. (\ref{eqMold}) the lower bound
$Q\ge - 3/4$ is easily obtained,  and the minimum
$Q_{\mbox{min}}=-3/4$
is obtained when $\Omega_{
\mbox{min}} = \Gamma/ \sqrt{2}$.

\section{Model and Generating Function Formalism} 

 Let $N$ be  the random  number of photons emitted by a single molecule
 source in a time interval $(0,t)$, and 
 $P_N(t)$ is the probability of $N$ emission events. 
The information
about the photon statistics is contained in
the moment generating function \cite{Brown}
\begin{equation}
2 { \cal Y } (s) \equiv
 \sum_{N=0}^\infty s^N P_N(t)
\label{eq02}
\end{equation}
which yields the moments of $N$
\begin{equation}
\begin{array}{c c}
\overline{N(t) } = 2  {\cal Y}' (1) \ \ \ \ \
\overline{N^2(t)} = 2 {\cal Y}'' (1)
+ 2 {\cal Y}' (1),
\end{array}
\label{eqA02}
\end{equation}
with which the $Q$ parameter can in principle be obtained.
In Eq. (\ref{eqA02}),
and in what follows,
we use the notation
\begin{equation}
{\partial \over \partial s} g(s)|_{s=1} \equiv g'(1)
\end{equation}
and similarly for second order derivatives with respect to
$s$.
The over-line in Eq. (\ref{eqA02}) describes an
average over the process of photon emission,
later we will consider a second type of average with respect
to the spectral diffusion process, which we will denote with
$\langle ... \rangle$.

The equations of motion for  the generating function was
given in \cite{Brown} and are called generalized optical Bloch
equation. 
For  a chromophore with single excited and
ground states, and interacting with a continuous wave laser field
\begin{equation}
\begin{array}{l}
 \dot{{ \cal U}}\left(s\right) = - {\Gamma \over 2} {\cal U}\left(s\right)
+ \delta_L(t) {\cal V}\left( s \right) \\
 \dot{{\cal V}}\left(s\right) = -
\delta_L(t) {\cal U}\left( s \right)-{\Gamma\over 2} {\cal V}\left( s \right)
- \Omega {\cal W} \left( s \right) \\
 \dot{{\cal W}}\left(s\right) =
\Omega {\cal V}\left( s \right)-{\Gamma\over 2} \left( 1 + s \right)
 {\cal W}\left( s \right)
- {\Gamma \over 2} \left( 1 + s\right) {\cal Y} \left( s \right) \\
\dot{{\cal Y} } \left( s \right) = - { \Gamma \over 2} \left( 1 - s \right)
{\cal W} \left( s \right) - { \Gamma \over 2} \left( 1 - s \right) {\cal Y} \left( s \right).
\end{array}
\label{eqA01}
\end{equation}
These equations are
exact within the rotating wave approximation and optical
Bloch equation formalism.  They yield the same type of information
on photon statistics
contained in the quantum jump approach to quantum optics
which is used in quantum Monte Carlo simulations
\cite{Plenio,Molmer}. 
In Eq. (\ref{eqA01}) $\Gamma$ is the spontaneous emission rate of
the electronic transition and
$\Omega$ is the Rabi frequency.
The time evolving detuning  is
\begin{equation}
\delta_L(t) = \omega_L - \omega_0 - \Delta\omega(t),
\end{equation}
where $\omega_L$ $(\omega_0)$ is the laser frequency
(the molecule's bare frequency),
and $\Delta\omega(t)$ is the stochastic spectral diffusion process.
In Eq. (\ref{eqA01}) it is 
assumed that the molecule
in its excited and ground state have no permanent dipole 
moments, hence the system is described only by the transition
dipole moment via the Rabi frequency.

The physical meaning of
${\cal U}(s)$, ${\cal V}(s)$,
and ${\cal W}(s)$ and their relation to
the standard Bloch equation was given in \cite{Brown}, 
some discussion on
this issue will follow Eq.
(\ref{eqMT}). For related work on the foundations
of these equations
see
\cite{Mukamel,Cook} and Ref. therein. 
Note that when $s \to 1$
the damping terms in Eq. (\ref{eqA01}) become small
[i.e. the $(1-s) \Gamma /2$ terms], hence relaxation of
the generalized Bloch equations in the important limit of
$s \to 1$ is slow.
                                                                                                                                                             
In what follows we will consider the
moments $\overline{N(t) }$, $\overline{N^2(t)}$. For this aim it
is useful to derive equations of motion for the
vector
$ z = $
\begin{equation}
\left\{
 {\cal U} (1), {\cal V}(1),{\cal W}(1),
{\cal Y}(1),{\cal U}'(1),
{\cal V}'(1),{\cal W}'(1),{\cal Y}'(1),{\cal Y}''(1) \right\}.
\end{equation}
 Taking the first and the second derivative of  Eq.
(\ref{eqA01}) with respect to $s$ and setting $s=1$, we find
\begin{equation}
\dot{z} = M(t) z
\label{eqmtz}
\end{equation}
where $M(t)$ is a $9\times 9$ matrix
$M(t) =$
\begin{equation}
\left(
\begin{array}{c c c c c c c c c}
-{\Gamma\over 2} & \delta_L (t) &  0  &  0  &  0  &  0  &  0  &  0  &  0 \\
-\delta_L(t) & -{\Gamma \over 2} & - \Omega&0  &  0  &  0  &  0  &  0  &  0 \\
0& \Omega & - \Gamma  & - \Gamma &    0  &  0  &  0  &  0  &  0  \\
0 & 0 & 0  & 0  &  0  &  0  &  0  &  0 & 0  \\
0 & 0 & 0  & 0  & - {\Gamma \over 2} & \delta_L (t) & 0  &  0 & 0 \\
0 & 0 & 0  & 0  & - \delta_L (t)  & - {\Gamma \over 2} &-\Omega & 0 & 0 \\
0 & 0 & -{\Gamma \over 2} &- {\Gamma \over 2} & 0 & \Omega &-\Gamma & -\Gamma & 0 \\
0 & 0 & {\Gamma \over 2} & {\Gamma \over 2} & 0 & 0 & 0 & 0 & 0 \\
0 & 0 &  0 & 0 & 0 & 0 & \Gamma & \Gamma & 0
\end{array}
\right)
\label{eqMT}
\end{equation}
The first three lines of $M(t)$ describe
the evolution of
${\cal U} (1), {\cal V}(1),{\cal W}(1)$, these are the standard
optical Bloch equations in the rotating wave approximation \cite{Wolf}.
These equations  yield ${\cal W}(1)$ which in turn gives the 
mean number of photons using Eq. 
(\ref{eqA02})
\begin{equation}
\dot{\overline{N}}(t) = \Gamma \left[ {\cal W}(1) + {1 \over 2} \right], 
\end{equation}
and us-usual ${\cal W}(1) + {1 \over 2} $ is the population 
in the excited state. 
The fourth line of $M(t)$ is zero, it yields
$\dot{\cal{Y}}(1)=0$, this equation describes the normalization
condition of the problem namely
${\cal Y}(1)=1/2$ for all times $t$ [to see this
use Eq. (\ref{eq02}) and $\sum_{N=0}^\infty P_N(t) = 1$].
The evolution of the other terms
${\cal U}'(1),
{\cal V}'(1),{\cal W}'(1),{\cal Y}''(1)$ are of
current  interest since they describe the fluctuation of the
photon emission process.
In particular using Eqs. (\ref{eqA02},\ref{eqMT}) 
\begin{equation}
{{\rm d} \over {\rm d} t}  
\overline{ N\left(t\right)\left[ N\left(t\right) - 1 \right] } = 
\Gamma\left[ \overline{N\left(t\right)}
+ 2 {\cal W}'(1) \right].
\end{equation}
Solutions of time dependent equations
like Eq.
(\ref{eqmtz}) are generally difficult to
obtain, a formal solution is given in terms of
the time ordering operator $T$,
$z(t) = T \exp [\int_0^t M(t) {\rm d} t] z(0)$.
                                                                                                                                                             
Eq. (\ref{eqmtz}) yields a general method for the
calculation of $Q$ for a single molecule undergoing
a spectral diffusion process.
The aim of this paper is to obtain an exact solution
of the problem for an important stochastic process
used by Kubo and Anderson 
\cite{AK,Kubo1,Kubo2} to investigate characteristic
behaviors of line shapes.
We assume $\Delta \omega(t) = \nu h(t)$ where $\nu$ describe
frequency shifts, and $h(t)$ describes a random telegraph process:
$h(t)=1$ or $h(t)=-1$. The transition rate between state
up (+) and state down (-) and vice versa is $R$.
This dichotomic process is sometimes called
the Kubo-Anderson process \cite{Kampen}. 
It was used to describe
generic
behaviors of line shapes 
\cite{Geva,ADV,Berne,Kampen},
here our aim is
to calculate $Q$ describing the line shape fluctuations.
                                                                                                                                                             
We use Burshtein's method \cite{Burshtein,Shore}
of marginal averages,
to solve the stochastic differential matrix equation
(\ref{eqmtz}). The method yields the average
$\langle z \rangle$ with respect to the stochastic
process. We will calculate $\langle z \rangle$ in the limit
of long times, and then obtain the steady state behavior
of  the line shape
\begin{equation}
I\left( \omega_L \right)=\lim_{t \to \infty} {{\rm d} \over {\rm d} t}\langle \overline{N} (t) \rangle
\end{equation}
and the  $Q$ parameter.
Let $\langle z \rangle_{\pm}$ be the average
of $z(t)$ under the condition that at time
$t$ the value of $h(t) = \pm 1$ respectively.
 $\langle z \rangle_{\pm}$ are called marginal averages,
the complete average is $\langle z \rangle = \langle z \rangle_{+}
+ \langle z \rangle_{-} $.
The equation of motion for the marginal averages is
an $18 \times 18$ matrix equation
\begin{equation}
\left(
\begin{array}{l}
\langle \dot{z}_{+} \rangle \\
\langle \dot{z}_{-} \rangle
\end{array}
\right)=
\left(
\begin{array}{c c}
M_{+} - R I & R I  \\
RI           & M_{-} - R I  \\
\end{array}
\right)
\left(
\begin{array}{l}
\langle z_{+} \rangle \\
\langle z_{-} \rangle
\end{array}
\right).
\label{eqMAT18}
\end{equation}
In Eq. (\ref{eqMAT18})
the matrix $M_{\pm}$ is identical to matrix $M$
in Eq.
(\ref{eqMT})
when $\delta_L(t)$ is replaced by
$\delta_L ^{\pm} =\omega_L - \omega_0 \mp \nu$,
 and $I$ is a $9\times 9$ identity matrix.
 In the next subsection we obtain the long time solution of Eq. 
(\ref{eqMAT18}), the reader not interested in the mathematical
details
may skip to subsection \ref{secExact}, where the solution
for the line shape
and $Q$
is presented.

\section{Mathematical Derivation of Exact Solution}

In three main steps, we  
now find the long time behavior of the
marginal averages, with which the line and $Q$ are then obtained.\\
${\bf 1. }$ 
As mentioned, 
from normalization condition
we have ${\cal Y}(1)=1/2$ for all times. 
Eq. (\ref{eqMAT18}) yields the marginal averages
$\langle {\cal Y}(1) \rangle_{+} = \langle {\cal Y}(1) \rangle_{-} = 1/4$,
in the steady state. Inserting these identities
in Eq. 
 (\ref{eqMAT18}) we obtain an equation of motion for the vector
\begin{equation}
y \equiv \langle \left\{
{\cal U} (1)_{+} , {\cal V}(1)_{+} , {\cal W}(1)_{+},
{\cal U} (1)_{-} , {\cal V}(1)_{-} , {\cal W}(1)_{-}
\right\} \rangle
\end{equation}  
\begin{equation}
\dot{y} = A y + b_0
\label{eqstep1}
\end{equation}
where
$A=$
\begin{equation}
\left(
\begin{array}{c c c c c c }
-{\Gamma\over 2} - R  & \delta^{+} _L &  0  &  R  &  0  &  0   \\
-\delta^{+}  _L & -{\Gamma \over 2} -R  & - \Omega &0  &  R  &  0  \\
0& \Omega & - \Gamma - R   & 0  &    0  &  R   \\
R & 0 & 0  &  -{\Gamma \over 2} -R  &  \delta^{-} _L   &  0  \\
0 & R & 0  & - \delta^{-} _L & - {\Gamma \over 2} - R & -\Omega   \\
0 & 0 & R  & 0  & \Omega & - \Gamma - R
\end{array}
\right)
\end{equation}
and $b_0 = \left( 0,0, - \Gamma/4 , 0, 0, -\Gamma/4 \right)$. 
In the long time limit the solution of Eq. (\ref{eqstep1})
reaches a steady state (ss) given by
\begin{equation}
y^{ss} = - A^{-1} b_0,
\label{eqstep11}
\end{equation}
and $A^{-1}$ is the inverse of $A$.
From Eq. (\ref{eqstep11}) we find
$$ \langle {\cal W}^{ss}(1) \rangle_{+} = { \Gamma \over 4} \left(
A_{33} ^{-1} + A_{36} ^{-1} \right) ,  $$
\begin{equation}
\langle {\cal W}^{ss}(1) \rangle_{-} = { \Gamma \over 4} \left(
A_{63} ^{-1} + A_{66} ^{-1} \right) ,   
\end{equation}
where $A_{36} ^{-1} = A_{63} ^{-1}$, and $A_{ij} ^{-1}$ is the
$ij$ matrix element of $A^{ - 1}$, $i,j = 1, \cdots ,6$.
We note that $\langle { \cal W}^{ss} (1) \rangle_{\pm}$
yields the steady state marginal averages of the population 
difference between the excited and ground state. 
From Eq.
(\ref{eqA02})
we see that  
we need $2 \langle {\cal Y}'(1) \rangle$
to obtain the average number of photon emissions $\langle \overline{N} \rangle$.
We use Eq. 
(\ref{eqMAT18}) and show
$$ \langle {\cal Y}'\ ^{ss} (1) \rangle = 
\langle {\cal Y}'\ ^{ss} (1) \rangle_{+} +  
\langle {\cal Y}'\ ^{ss} (1) \rangle_{-} =  $$
\begin{equation}
{ \Gamma \over 2} \left( 
\langle {\cal W}^{ss} \left( 1 \right) \rangle_{+} +
\langle {\cal W}^{ss} \left( 1 \right) \rangle_{-} + { 1  \over 2} \right) t
\label{eqyd}
\end{equation}
and
\begin{equation}  
\langle {\cal Y}'\ ^{ss} \left( 1 \right) \rangle_{+} -
\langle {\cal Y}'\  ^{ss} \left( 1 \right) \rangle_{-} =
{ \Gamma \left( \langle {\cal W}^{ss} \left( 1 \right) \rangle_{+}
-       \langle {\cal W}^{ss} \left( 1 \right) \rangle_{-} \right) \over 4 R} .
\end{equation}
The average number of emitted photon, in the long time
limit is
(\ref{eqA02},
\ref{eqyd})
\begin{equation}
\langle \overline{N} \rangle = 
 \Gamma \left( 
\langle {\cal W}^{ss} \left( 1 \right) \rangle_{+} +
\langle {\cal W}^{ss} \left( 1 \right) \rangle_{-} + {1 \over 2} \right) t,
\end{equation}  
namely 
$\langle \overline{N} \rangle $  is proportional to 
the steady state occupation in the excited state. 
The line shape is 
\begin{equation}
I\left( \omega_L \right)  = 
 {I_{+} + I_{-} \over 2}
\end{equation}
where
\begin{equation}
I_{\pm} = 2 \Gamma \left[ \langle {\cal W}^{ss}  (1)\rangle_{\pm} + {1 \over 4} 
\right].
\end{equation}

${\bf 2. }$ We use the solutions
obtained in previous step to obtain inhomogeneous equations
for
$x= \langle \{
 {\cal U}'(1)_{+}, {\cal V}'(1)_{+},{\cal W}'(1)_{+},
 {\cal U}'(1)_{-}, {\cal V}'(1)_{-},{\cal W}'(1)_{-}
\} \rangle$, 
\begin{equation}
\dot{x} = A x + b(t)
\label{eqXAB}
\end{equation}
where
%
%\begin{equation}
%A=\left(
%\begin{array}{c c c c c c }
%-{\Gamma\over 2} - R  & \delta^{+} _L &  0  &  R  &  0  &  0   \\
%-\delta^{+}  _L & -{\Gamma \over 2} -R  & - \Omega &0  &  R  &  0  \\
%0& \Omega & - \Gamma - R   & 0  &    0  &  R   \\
%R & 0 & 0  &  -{\Gamma \over 2} -R  &  \delta^{-} _L   &  0  \\
%0 & R & 0  & - \delta^{-} _L & - {\Gamma \over 2} - R & -\Omega   \\
%0 & 0 & R  & 0  & \Omega & - \Gamma - R
%\end{array}
%\right)
%\end{equation}
and $b(t) = ( 0, 0, b_{+}(t) , 0, 0, b_{-}(t) )$ with
\begin{widetext}
\begin{equation}
 b_{\pm}(t) = -{\Gamma \over 8} - {\Gamma \over 2} 
 \langle{\cal W}^{ss}(1)\rangle_{\pm}-
{\Gamma^2 t \over 8} \left\{
1 + 2 \left[ \langle{\cal W}^{ss}(1)\rangle_{+} +\langle {\cal W}^{ss}(1)\rangle_{-} \right] \right\}
 \pm { \Gamma^2 \over 8 R } \left[ \langle {\cal W}^{ss}(1)\rangle_{-} - 
 \langle{\cal W}^{ss}(1)\rangle_{+} \right].
\label{eqBB}
\end{equation}
In the long time limit we obtain 
\begin{equation}
x(t) \sim A^{ - 1} c_0 + \left( A^{ - 1} t + A^{ - 1} A^{-1}\right) c_1
\end{equation}
where $c_0$ and $c_1$ are column vectors
\begin{equation}
c_0 = \left( 0, 0 , c_{+} , 0, 0, c_{-} \right), \ \ \
c_1 = \left( 0, 0 , {\Gamma I\left( \omega_L\right)\over   4} , 0, 0,
{ \Gamma I \left( \omega_L \right) \over 4} \right),
\end{equation}
with
\begin{equation}
c_{\pm} = {\Gamma \over 2} \left[
\langle {\cal W}^{ss}\left( 1 \right) \rangle_{\pm} + {1 \over 4} 
\right] \mp {\Gamma^2 \over 8 R} \left[
\langle {\cal W}^{ss} \left(1 \right)  \rangle_{-} -
\langle {\cal W}^{ss} \left(1 \right)  \rangle_{+} \right]
\end{equation}
we therefore obtain
\begin{equation}
\langle {\cal W}'\ ^{ss} \left( 1 \right) \rangle_{+} =
A_{33}^{ - 1} c_{+} + A_{3 6} ^{ - 1} c_{-} +
I(\omega_L  ) \langle {\cal W}^{ss} \left( 1 \right) \rangle_{+} t+
{ \Gamma I \left( \omega_L \right) \over 4} \left[
\left( A^{-1} A^{ -1} \right)_{33} +
\left( A^{-1} A^{ -1} \right)_{36}\right] 
\end{equation}
\begin{equation}
\langle {\cal W}'\ ^{ss} \left( 1 \right) \rangle_{-} =
A_{63}^{ - 1} c_{+} + A_{6 6} ^{ - 1} c_{-} +
I(\omega_L ) \langle {\cal W}^{ss} \left( 1 \right) \rangle_{-} t+
{ \Gamma I \left( \omega_L  \right) \over 4} \left[
\left( A^{-1} A^{ -1} \right)_{63} +
\left( A^{-1} A^{ -1} \right)_{66}\right]. 
\end{equation}

${\bf 3.}$ From Eq. 
(\ref{eqA02}),
$\langle \overline{N^2(t)} \rangle  = \langle 2 {\cal Y}'' (1) \rangle
+ \langle  2 {\cal Y}' (1) \rangle$. In steady state we have
\begin{equation}
\langle {\cal Y}{''} ^{ss} \left( 1 \right) \rangle =  
\langle {\cal Y}{''} ^{ss} \left( 1 \right) \rangle_{+} +   
\langle {\cal Y}{''} ^{ss} \left( 1 \right) \rangle_{-}.   
\end{equation}
From Eq. 
(\ref{eqMAT18})  one can show that 
in the long time limit
$$ \langle {\cal Y}''\ ^{ss} (1) \rangle  \sim
4 \left[ \langle {\cal W}^{ss} (1) \rangle_{+} c_{+} +
\langle {\cal W}^{ss} (1) \rangle_{-} c_{-} \right] t  + $$
$$ +{\Gamma^2 I(\omega_L) \over 4} 
\left[ \left( A^{-1} A^{-1} \right)_{33} + \left( A^{-1} A^{-1} \right)_{36}
+ \left( A^{-1} A^{-1} \right)_{63} + \left( A^{-1} A^{-1} \right)_{66}
\right] t $$
\begin{equation}
+ { \Gamma I \left( \omega_L\right) \over 2} \left[
\langle {\cal W}^{ss} (1) \rangle_{+} +
\langle {\cal W}^{ss} (1) \rangle_{-} + {1 \over 2} \right]t^2.
\label{eqydd}
\end{equation}
 Finally we obtain the $Q$ parameter using:
\begin{equation}
Q=  { \langle  {\cal Y}''(1)^{{\rm ss}} \rangle -2 \langle
 {\cal Y}'(1)^{{\rm ss}} \rangle^2 \over
 \langle {\cal Y}'(1)^{{\rm ss}} \rangle}.
\label{EqDef}
\end{equation}
Using Eqs. (\ref{eqyd},\ref{eqydd},\ref{EqDef}) 
we obtain the main result of this manuscript
\begin{equation}
 Q =
 {\Gamma^2 \over 2} \sum_{i=3,6} \sum_{j=3,6} \left( A^{-1}A^{-1} \right)_{ij}
+ {\Gamma^2 \over R I (\omega_L) }\left[ \langle {\cal W}^{ss}(1)\rangle_{+} - \langle{\cal W}^{ss}(1)\rangle_{-} \right]^2 +
{\Gamma \over I\left( \omega_L \right)}
\sum_{k=\pm}  \left\{ \langle {\cal W}^{ss}(1)\rangle_{k} \left[ 1 +
4 \langle {\cal W}^{ss}(1)\rangle_{k} \right] \right\},
\label{eqMain}
\end{equation}
which is valid when measurement time $t \to \infty$.
The $Q$ parameter in Eq. (\ref{eqMain}) is expressed in terms of
$A^{-1}$.
To obtain the solution in terms of
the original parameters of the problem 
$R,\nu, \omega_L, \omega_0, \Omega,\Gamma$,
we found analytical expressions for 
$A^{-1}$ using Mathematica. The formula for the $Q$ parameter is
given in the following subsection.

\end{widetext}

\subsection{Exact Solution }
\label{secExact}

Without loss of generality we set $\omega_0=0$, hence 
$\omega_L$ is the detuning. We find
\begin{equation}
Q={\mbox{Numerator}\left[Q\right] \over \mbox{Denominator}\left[Q\right] }, 
\label{eqQ}
\end{equation}
$$\mbox{Numerator}\left[Q\right] = $$
$   -2 \Gamma \Omega^2 (R (\Gamma + 4 R) (3 \Gamma^3 - 4 \Gamma \nu^2 + 24 \Gamma^2 R + 
   16 \nu^2 R + 48 \Gamma R^2) (\Gamma^3 + 6 \Gamma^2 R + 8 \nu^2 R +
    2 \Gamma (2 \nu^2 + \Omega^2 + 4 R^2))^3 + 
   8 (4 \Gamma^{11} R - 4 \Gamma^{10} (\nu^2 - 21 R^2) - 3072 \Gamma \nu^4 R^5 
    (4 \nu^2 - \Omega^2 - 16 R^2) 
 - 4096 \nu^4 R^6 (2 \nu^2 - \Omega^2 - 4 R^2) +
   15 \Gamma^9 R (\Omega^2 + 48 R^2) 
    - 8 \Gamma^8 (4 \nu^4 - 27 \Omega^2 R^2 -
    392 R^4 + 2 \nu^2(\Omega^2 - 48 R^2)) - 
   4 \Gamma^5 R (128 \nu^6 + \Omega^6 + 48 \Omega^4 R^2 + 1344 \Omega^2 R^4 + 
     6400 R^6 + 12 \nu^4 (11 \Omega^2 - 272 R^2) +
     4 \nu^2 (9 \Omega^4 - 302 \Omega^2 R^2 - 5072 R^4)) - 
    64 \Gamma^3 R^3 (48 \nu^6 + 3 \nu^4 (11 \Omega^2 - 448 R^2) +
     3 (\Omega^2 + 4 R^2)^2 (\Omega^2 + 16 R^2) + 
     2 \nu^2 (5 \Omega^4 - 80 \Omega^2 R^2 - 544 R^4)) -
     256 \Gamma^2 R^4 (28 \nu^6 + \nu^4 (3 \Omega^2 - 332 R^2) + 
     (\Omega^2 + 4 R^2)^3 + 2 \nu^2 (\Omega^4 - 4 \Omega^2 R^2 - 32 R^4)) +
     4 \Gamma^7 R (-48 \nu^4 - 2 \nu^2 (9 \Omega^2 - 952 R^2) + 
     3 (\Omega^4 + 89 \Omega^2 R^2 + 544 R^4)) -
     8 \Gamma^6 (8 \nu^6 + 8 \nu^4 (\Omega^2 - 17 R^2) + 
     2 \nu^2 (\Omega^4 - 38 \Omega^2 R^2 - 2116 R^4) -
     3 (3 \Omega^4 R^2 + 58 \Omega^2 R^4 + 40 R^6)) - 
     16 \Gamma^4 R^2 (96 \nu^6 + 96 \nu^4(\Omega^2 - 31 R^2) +
     \nu^2(26 \Omega^4 - 720 \Omega^2 R^2 - 6656 R^4) + 
     3 (\Omega^6 + 44 \Omega^4 R^2 + 448 \Omega^2 R^4 + 1152 R^6))) \omega_L^2 +
     32 (\Gamma + 2 R)(3 \Gamma^8 R + 
     512 \Gamma \nu^4 R^4 + 512 \nu^4 R^5 -
8 \Gamma^7(\nu^2 - 3 R^2) + \Gamma^6(-32\nu^2 R + 3 R(\Omega^2 - 4 R^2)) - 
     32 \Gamma^2 R^3 (-14 \nu^4 + 6 \nu^2 \Omega^2 + 3 (\Omega^2 + 4 R^2)^2) +
     2 \Gamma^5 (16 \nu^4 - 8 \nu^2 (\Omega^2 - 3 R^2) - 
     3 R^2 (5 \Omega^2 + 104 R^2)) - 8 \Gamma^3 R^2 
     (-56 \nu^4 + \nu^2 (29 \Omega^2 - 40 R^2) + 
     6 (\Omega^4 + 20 \Omega^2 R^2 + 64 R^4)) -
     2 \Gamma^4 R (-104 \nu^4 + 10 \nu^2 (5 \Omega^2 - 16 R^2) + 
     3 (\Omega^4 + 60 \Omega^2 R^2 + 368 R^4))) \omega_L^4 -
     128 \Gamma^2 (\Gamma + 2 R)^2(4\Gamma^2(\nu^2 + 3 R^2) + 
     \Gamma R (16 \nu^2 + 3 \Omega^2 + 48 R^2) +
     4 R^2 (4 \nu^2 + 3 (\Omega^2 + 4 R^2))) \omega_L^6 - 
     256 \Gamma^2 R (\Gamma + 2 R)^3 \omega_L^8) $
$$\mbox{Denominator}\left[Q\right] = $$
$   R((\Gamma + 4 R) (\Gamma^3 + 6 \Gamma^2 R + 8 \nu^2 R +
   2 \Gamma (2 \nu^2 + \Omega^2 + 4 R^2)) + 4 \Gamma(\Gamma + 2 R) \omega_L^2)
   (\Gamma^6 + 10 \Gamma^5 R + 64 \nu^2 \Omega^2 R^2 +
    4 \Gamma^4 (2 \nu^2 + \Omega^2 + 8 R^2) +
    4 \Gamma^3 R (12 \nu^2 + 7 \Omega^2 + 8 R^2) +
    16 \Gamma R (2 \nu^4 + 3 \nu^2 \Omega^2 + \Omega^4 + 4 \Omega^2 R^2) +
    4 \Gamma^2 (4 \nu^4 + \Omega^4 + 16 \Omega^2 R^2 + 4 \nu^2 (\Omega^2 + 4 R^2)) +
    8 \Gamma (\Gamma^3 + 6 \Gamma^2 R - 8 \nu^2 R + 6 \Omega^2 R + 16 R^3 +
    2 \Gamma (-2\nu^2 + \Omega^2 + 8 R^2)) \omega_L^2 +
    16 \Gamma (\Gamma + 2 R) \omega_L^4)^2. $\\
The line shape is
 \begin{equation}
 I \left(\omega_L \right)  = 
{
\mbox{Numerator}[I\left(\omega_L \right)] \over 
                                  \mbox{Denominator}[I\left(\omega_L \right)] } ,
\label{eqLS}
\end{equation}
$$\mbox{Numerator}\left[I\left(\omega_L \right)\right] = $$
$   \Gamma \Omega^2 ((\Gamma + 4 R) (\Gamma^3 + 6 \Gamma^2 R + 8 \nu^2 R +
2 \Gamma (2 \nu^2 + \Omega^2 + 4 R^2)) + 4 \Gamma (\Gamma + 2 R) \omega_L^2), $
$$\mbox{Denominator}\left[I\left(\omega_L \right)\right] = $$
$ \Gamma^6 + 10 \Gamma^5 R + 64 \nu^2 \Omega^2 R^2 +
4 \Gamma^4 (2 \nu^2 + \Omega^2 + 8 R^2 ) +
  4 \Gamma^3 R (12 \nu^2 + 7 \Omega^2 + 8 R^2) +
  16 \Gamma R (2 \nu^4 + 3 \nu^2 \Omega^2 + \Omega^4 + 4 \Omega^2 R^2) +
  4 \Gamma^2 (4 \nu^4 + \Omega^4 + 16 \Omega^2 R^2 + 4 \nu^2 (\Omega^2 + 4 R^2)) +
  8 \Gamma (\Gamma^3 + 6 \Gamma^2 R - 8 \nu^2 R + 6 \Omega^2 R + 16 R^3 +
  2 \Gamma (-2 \nu^2 + \Omega^2 + 8 R^2)) \omega_L^2 +
  16 \Gamma (\Gamma + 2 R) \omega_L^4. $

\subsubsection{Exact Solution for Zero Detuning}

When the detuning is zero we find 
\begin{equation}
Q\left(\omega_L = 0\right) = - { 2 \Gamma \Omega^2 
\left[ 4 \nu^2 \left( - \Gamma + 4 R \right) + 
3 \Gamma \left( \Gamma + 4 R \right)^2 \right] \over
\left[ 4 \Gamma \nu^2 + \left( \Gamma^2 + 2 \Omega^2 \right) 
\left( \Gamma + 4 R \right) \right]^2 }.
\label{eqZT}
\end{equation}
Using Eq. (\ref{eqZT}) the 
lower bound $Q\left(\omega_L = 0\right)>-3/4$ is obtained.
The absolute minimum of $Q$, i.e.  $Q=-3/4$ 
is found when $\nu=0,\Omega = \Gamma/ \sqrt{2}$,
or when $R \to \infty,\Omega=\Gamma/ \sqrt{2}$.
Namely the absolute minimum is found
for a molecule whose absorption frequency is
fixed, or for a very fast spectral modulation.

{\bf Remark:}
Eq. (\ref{eqZT}) indicates a transition from sub-Poissonian
statistics $(Q<0)$ 
to super-Poissonian statistics $(Q>0)$
when
$4 \nu^2 = 3 ( \Gamma + 4 R)^2 / (  \Gamma - 4 R) $. 
If $R> \Gamma/4$, i.e. if the bath is fast compared with
the radiative life time, we find sub-Poissonian behavior for all values of
$\nu$ and $\Omega$. 

\begin{figure}
\begin{center}
\epsfxsize=70mm
%\epsfbox{figb2.eps}
\epsfbox{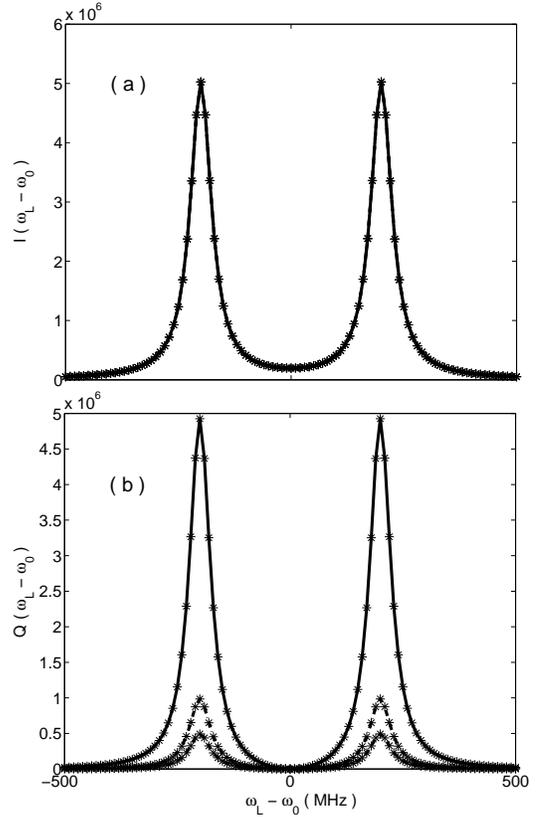}
\end{center}
\caption
{
 The line shape $I(\omega_L -\omega_0)$ (a) and
the $Q(\omega_L - \omega_0)$ parameter (b) 
are calculated with the exact solution  Eqs.
(\ref{eqLS}) and Eq. (\ref{eqQ}) respectively. In this slow modulation
strong coupling limit 
 $ R << \Gamma \simeq \Omega << \nu $,  
both $Q$ and $I$ exhibit
splitting behavior with two peaks 
centered at $ \omega_L - \omega_0 = \pm \nu $. 
The $Q$ parameter exhibits super-Poissonian statistics $Q> 0$.
The $*$ asterisks are approximate results
Eq. (\ref{eqsl01}) for the line shape,
and Eq. (\ref{eqQ1})
for $Q$. 
 Parameters are 
$\Gamma = 40$ MHz, $\nu = 5 \Gamma$, $ \Omega = \Gamma / \sqrt{2} $,
and $ R = 1 $ Hz (solid curve), $ 5 $ Hz (dashed-dot curve), and  $ 25 $ 
Hz (dashed curve). 
}
\label{fig1}
\end{figure}

\section{The Physical Behaviors of the Exact Solution in Limiting Cases}

The $Q$ parameter is a function of two control
parameters $\omega_{L}$ and $\Omega$, and three model parameters
$\Gamma, R, \nu$. The classification of different types of
physical behaviors, based on the relative magnitude of 
the parameters is investigated in this section.
The limiting behaviors of $Q$ are obtained from the exact
solution using Mathematica.

\subsection{Slow Modulation Regime: $ R << \nu, \Gamma, \Omega $}

 In the slow modulation regime, the bath fluctuation process
(R) is very slow compared with 
the radiative decay rate 
($\Gamma$), frequency fluctuation amplitude ($\nu$) and the Rabi frequency
($\Omega$).
This case is similar to situations in many single molecule
experiments, for example single molecules in low temperature
glasses.

The exact solution can be simplified in the limit 
$R \rightarrow 0 $,  using
Eq. (\ref{eqLS}) 
\begin{equation}
\lim_{R \to 0} I\left( \omega_L \right) = 
{ I_{+}\left(\omega_L \right) + I_{-}\left( \omega_L \right) \over 2},
\label{eqsl01}
\end{equation}
and
\begin{equation}
I_{\pm}\left(\omega_L \right)  = 
{\Gamma \Omega^2 \over \left( \Gamma^2 + 2 \Omega^2 + 4 \left( \omega_L \mp \nu \right)^2 \right)}.
\label{eqLSpm1}
\end{equation}
The line is a sum of two Lorentzians centered on
$\pm \nu$,  namely it exhibits splitting behavior
when $\nu >> \Gamma,\Omega$. 
Using Eq. (\ref{eqQ}) we find a simple super-Poissonian behavior 
for $Q$ in the slow modulation limit $R \to 0$
\begin{equation}
Q\sim Q_{\mbox{slow}}   \equiv {\left( I_{+} - I_{-} \right)^2 \over 4 R I},
\label{eqQ1}
\end{equation}
and since $R \to 0$,  $Q$ may become very large
(e.g. $Q=5*10^6$ in Fig.  \ref{fig1}).

 A simple 
picture can be used to understand these results. In the slow
modulation limit the
molecule jumps between two states $+$ and $-$, the
time between successive jumps is very long, in such a way
that many photons are emitted between jump events. 
In each of these two states the molecule emits photons
at a rate $I_{\pm} (\omega_L) $, 
Eq. (\ref{eqLSpm1}).
These rates are determined by the  familiar steady state
solutions of the optical Bloch
equation, for a two level atom with the absorption frequency
$\omega_0 \pm \nu$ fixed in time \cite{Wolf}.  
In this slow limit
the random number of emitted photons,
in time interval $(0,t)$, is
$N_{\mbox{slow}} = \int_{0} ^t I(t) {\rm d} t$,
and $I(t)$ is a stochastic intensity that jumps between two
states $I_{\pm}(\omega_L)$ with the rate $R$. 
Using this simple random walk picture it is straightforward to derive Eqs.
(\ref{eqsl01},
\ref{eqQ1}).  For mathematical details 
see Ref. \cite{ADV} who considered a similar slow modulation limit 
which is valid only for weak Rabi frequency.

\begin{figure}
\begin{center}
\epsfxsize=70mm
%\epsfbox{figb4.eps}
\epsfbox{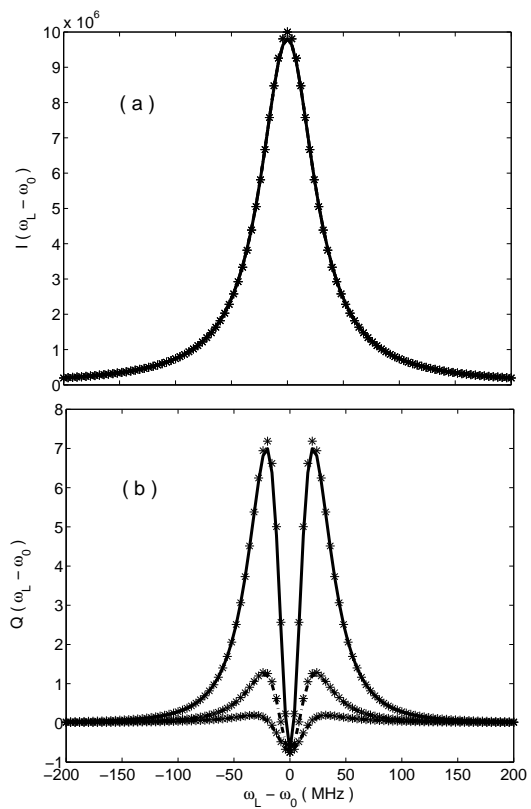}
\end{center}
\caption
{
Same as Fig. \ref{fig1} for the slow modulation
weak coupling case
$ R << \nu << \Gamma \simeq \Omega$.
We see that the line $I$ is Lorentzian in shape while
the $Q$ parameter exhibits splitting. And unlike the
line shape $Q$ depends on $R$ in this limit.  
In the vicinity of
zero detuning $Q$ exhibits a sub-Poissonian behavior ($Q < 0$). Far
from zero detuning  
the slow spectral diffusion process controls the behavior of $Q$
and then $Q>0$.
The $*$ asterisk are the approximate Eqs. 
for $Q$ 
Eq.
(\ref{eqQ2}) and $I$ Eq. 
(\ref{eqS02}).
The parameters are  
$\Gamma = 40$ MHz, $\nu = \Gamma / 10 $, $ \Omega = \Gamma / \sqrt{2} $,
and $ R = \Gamma / 2500 $ (solid curve), $ \Gamma /500 $ (dashed-dot curve), and $ \Gamma / 100 $
(dashed curve).
}
\label{fig2}
\end{figure}

\subsubsection{ $R<< \Gamma \simeq \Omega << \nu$}

 Within the slow modulation limit we distinguish between two cases.
The case $R << \Gamma \simeq \Omega << \nu$ is called the slow modulation
strong
coupling limit. In this case the line
and $Q$ have two well separated peaks
and the broadening of the two peaks due to
the finite life time, and power of the laser field is small  
compared with the frequency shifts (see Fig. \ref{fig1}).
From Fig. \ref{fig1}, and as expected from Eqs.
(\ref{eqsl01},
\ref{eqQ1}),
$Q$ decreases when $R$ is
increased, while $I(\omega_L)$ is independent of $R$.
Thus it is $Q$ not $I\left( \omega_L \right)$ that yields information
on the dynamics.
In Fig. \ref{fig1} the agreement between the exact  solution
and the approximation 
Eq. (\ref{eqQ1}) is good. \\

\subsubsection{$ R << \nu << \Gamma \simeq \Omega $}

The limit $R << \nu <<  \Gamma \simeq \Omega$ is called the weak
coupling slow modulation limit. In this case the two peaks of the line,
discussed in the previous sub-section,
are overlapping and the line is approximated by
\begin{equation}
I\left(\omega_L \right)  \sim  {\Gamma \Omega^2 \over  \Gamma^2 + 2 \Omega^2
 + 4 \omega_L^2}.
\label{eqS02}
\end{equation}
This result is exact when  $\nu = 0$, for arbitrary $R$, 

The behavior of $Q$ is demonstrated in Fig. \ref{fig2},
where we observe both super-Poissonian and sub-Poissonian
behaviors.
In 
the slow modulation weak coupling limit,
we must distinguish between the cases of
large and small detuning. First note that
according to Eq. 
(\ref{eqQ1}) 
when the detuning is zero we find 
$Q=0$, namely the leading $1/R$ term in our asymptotic expansion
vanishes. We must therefore consider the higher order terms
in our asymptotic expansion of Eq. 
(\ref{eqQ})
and we find
\begin{equation}
 Q \sim Q_{\mbox{slow}}  + Q_{\mbox{M}}
\label{eqQapp}
\end{equation}
where
\begin{equation}
 Q_{\mbox{M}}
=       - { 2 \Omega^2 \left( 3 \Gamma^2 - 4 \omega_L ^2 \right) \over \left( \Gamma^2 + 2 \Omega^2 + 4 \omega_L^2 \right)^2}.
\label{eqMandel}
\end{equation}
Eq. (\ref{eqQapp}) has a simple meaning, the first term is a contribution
to $Q$ from spectral diffusion, which is identical to
Eq. (\ref{eqQ1}). 
The second term $Q_{\mbox{M}}$ is identical to the result obtained by
Mandel, for the $Q$ parameter in the absence of
spectral diffusion \cite{Mandel1}, and $Q_M<0$ provided
that the detuning is not too large.
The second term is dominating over the first when the
detuning is small, and for zero detuning 
we obtain 
in Eq. (\ref{eqQapp})
sub-Poissonian statistics. 
More explicitly, we Taylor expand
(\ref{eqLSpm1}) using  $\nu$ as a small parameter, and obtain
for the slow modulation weak coupling limit
\begin{equation}
  Q   \simeq {64 \Gamma \Omega^2 \omega_L^2 \nu^2 \over R  \left( \Gamma^2 + 2 \Omega^2 + 4 \omega_L^2 \right)^3}   
- {2 \Omega^2 \left( 3 \Gamma^2 - 4 \omega_L ^2 \right)  \over \left( \Gamma^2 + 2 \Omega^2 + 4 \omega_L^2 \right)^2}.
\label{eqQ2}
\end{equation}
One may say that for zero detuning, the
molecule behaves as if its absorption frequency
is fixed.  

 To conclude, we see that in the slow modulation limit
$Q$ is a sum of two additive contributions: i) a part
related to spectral diffusion $Q_{\mbox{slow}}$ and ii) and a part
related to  quantum fluctuations,
i.e. $Q_M$. Such a behavior was
very recently discussed  in \cite{Brown2} (see also \cite{ADV}
for related discussion). The quantum fluctuations
are however bounded from above and below $-3/4<Q_{M}<0$, while the
contribution from spectral diffusion is not 
$ 0< Q_{\mbox{slow}} <\infty$. Hence detection of
the quantum fluctuations is possible only
when $Q_{\mbox{slow}} $ is small, which for
our case implies zero detuning and
weak coupling limit.

\begin{figure}
\begin{center}
\epsfxsize=70mm
%\epsfbox{figb6.eps}
\epsfbox{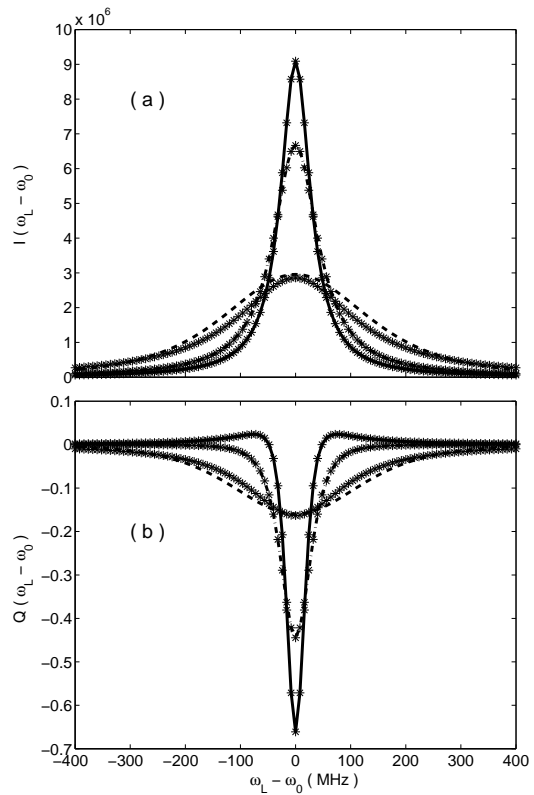}
\end{center}
\caption
{  Same as Fig. \ref{fig1} for fast modulation
$ R >> \nu >> \Omega = \Gamma/\sqrt{2}$.
We see quantum fluctuations namely $Q<0$. 
Both for the line shape and for $Q$ we have motional narrowing behavior,
as the spectral diffusion process
becomes faster $Q$ and $I$ narrow. 
When $R \to  \infty$, $Q$ approaches
Mandel's formula Eq. 
(\ref{eqMandel}).
In the same limit the width of the line shape
is determined only by its natural width,
namely by the inverse radiative life time of the excited state
(provided that the Rabi frequency is weak).
The $*$ asterisk are the approximations
for $Q$ 
Eq.
(\ref{eqQ3}) and $I(\omega_L)$ Eq. 
(\ref{eqLS3}).
The parameters are 
$\Gamma = 40$ MHz, $\nu =  5 \Gamma $, $ \Omega = \Gamma / \sqrt{2} $,
and $ R = 5 \Gamma $ (dashed curve), $ 25 \Gamma $ (dashed-dot curve), and $ 125 \Gamma $
(solid curve).
}
\label{fig3}
\end{figure}

\subsection{Fast Modulation Regime: $ R >> \nu, \Gamma, \Omega $}

 We now consider the fast modulation limit. If we take
 $R \to \infty$ keeping $\Omega$, $\Gamma$, and $\nu$ fixed
 we find from Eq. (\ref{eqQ})
 \begin{equation}
\lim_{R \to \infty} Q= Q_{\mbox{M}}
\end{equation}
given in Eq. (\ref{eqMandel}).
Hence in this limit we find a sub-Poissonian behavior, 
provided that the detuning $\omega_L$ is not too large.
The line shape is identical to the expression on the
right hand side of Eq. 
(\ref{eqS02}). This behavior is expected, when the spectral
diffusion is very fast 
the emitting single molecule cannot respond
to the stochastic fluctuations.

\begin{figure}
\begin{center}
\epsfxsize=70mm
%\epsfbox{figb14.eps}
\epsfbox{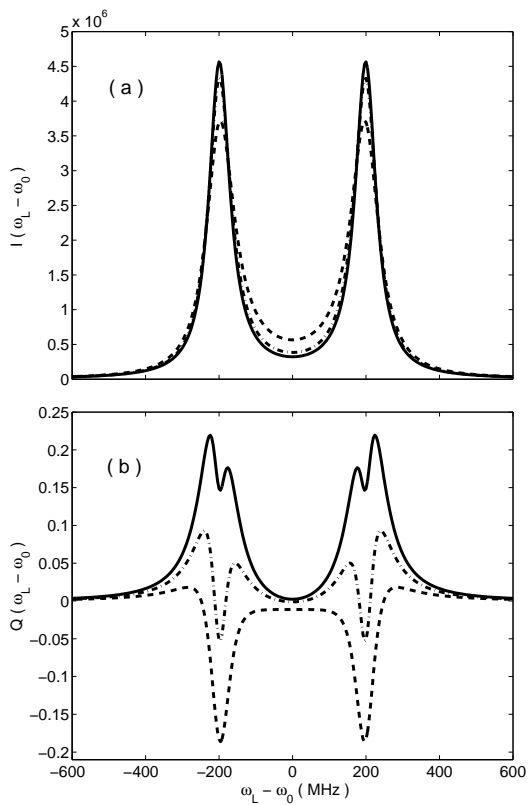}
\end{center}
\caption
{We show the transition of the $Q$ parameter from
sub-Poissonian behavior, to super Poissonian behavior as the rate $R$
is changed.
Note that $Q$ exhibits a non-trivial behavior:
it may have four peaks. Such a behavior is limited to a small
regime of parameters when $R \simeq \Gamma\simeq \Omega$ and was not
observed in other limits of the problem. 
The parameters are $\Gamma = 40$ MHz, $\nu = 5 \Gamma $, $ \Omega = \Gamma / \sqrt{2} $,
and $ R = \Gamma / 6 $ (solid curve), $ \Gamma /4 $ (dashed-dot curve), and $ \Gamma / 2 $
(dashed curve). 
}                                                                               \label{fig4}
\end{figure}

A more interesting case is to let $R \to \infty$
and $\nu \to \infty$
but keep, 
\begin{equation}
\Gamma_{{\rm SD}} \equiv { \nu^2 \over  R }
\end{equation}
finite.
We call this limit the fast modulation limit, 
using Eq. (\ref{eqLS})  the line shape 
is 
\begin{equation}
I_{\mbox{fast}} \left(\omega_L \right)  ={ \left( \Gamma + \Gamma_{{\rm SD}} \right ) \Omega^2 \over \left( \Gamma_{{\rm SD}} + \Gamma \right)^2
  + 2 \left( 1 + \Gamma_{{\rm SD}} / \Gamma \right) \Omega^2 + 4 \omega_L^2},
\label{eqLS3}
\end{equation} 
and when the Rabi frequency is small
\begin{equation}
I_{\mbox{ fast}}  \left(\omega_L \right)  \sim { \left( \Gamma + \Gamma_{{\rm SD}} \right ) \Omega^2 \over \left( \Gamma_{{\rm SD}} + \Gamma \right)^2
+ 4 \omega_L^2}.
\label{eqLS3a}
\end{equation} 
In this limit we have the well known effect of motional
narrowing: the width of the line is determine by
$\Gamma + \Gamma_{{\rm SD}} $ and as the process becomes faster
the line becomes narrower, namely $\Gamma_{{\rm SD}}$
decreases when $R$ is increased. 

The $Q$ parameter is obtained from Eq. (\ref{eqQ})
$$ Q_{\mbox{fast}}  = $$
\begin{equation}
- { 2 \Gamma \Omega^2 \left[ 3 \Gamma^3 + 5 \Gamma \Gamma_{{\rm SD}}^2 + \Gamma_{{\rm SD}}^3 - 4 \Gamma \omega_L\ ^2 
+ \Gamma_{{\rm SD}} \left( 7 \Gamma^2 + 4 \omega_L ^2 \right) \right]
\over
\left[ \Gamma^3 + \Gamma \Gamma_{{\rm SD}}\ ^2 + 2 \Gamma \Omega^2 
+ 2 \Gamma_{{\rm SD}} \left( \Gamma^2 + \Omega^2 \right) + 4 \Gamma \omega_L ^2 \right]^2 }.
\label{eqQ3}
\end{equation}
Thus in the fast modulation limit 
the photon statistics is sub-Poissonian provided that
the detuning is not too large. When $\Gamma_{SD} \to 0$ the
result for $Q$ reduces to Mandel's result Eq. (\ref{eqMandel}).
In Fig. \ref{fig3}, we show
the line shape and the $Q$ parameter, for three 
values of the jump rate (R) in the fast modulation regime 
while $\nu,\Omega,\Gamma$ are kept fix.
We see that as the stochastic spectral diffusion process
gets faster, both the line shape 
and the $Q$ parameter become narrow. Thus 
both $I(\omega_L)$ and 
$Q$ exhibit a motional narrowing effect. Also, 
as the stochastic
process gets faster, a stronger quantum behavior
is obtained, in the sense that the minimum of $Q$ 
decreases.

\subsection{Strong coupling limit $ \nu >> R , \Gamma, 
\Omega$}

To investigate the strong coupling limit  
we consider the value of $Q$ for $ \omega_L = \nu$,
and $ \nu >> \Gamma,R,\Omega$.
From Eq. (\ref{eqQ}) we obtain
\begin{widetext}
\begin{equation}
 \lim_{\nu \to \infty} Q_{\omega_L=\nu} = 
-{ \left( \Gamma + 2 R \right) \Gamma \Omega^2 
\left[ - \Gamma^3 + 16 \Gamma R^2 + 8 R^3 - 2 \Gamma \Omega^2 +
2 R \left( 2 \Gamma^2 + \Omega^2 \right) \right]
\over 2 R \left[ \Gamma^3 + 4 \Gamma R^2 + 2 \Gamma \Omega^2 + 2 R \left( 2\Gamma^2 + \Omega^2 \right) \right]^2 }.
\label{eqwwww}
\end{equation}
\end{widetext}
This
equation
exhibits both sub-Poissonian
and super--Poissonian behaviors.
When the process is very {\em slow}, namely $R \to 0$, we obtain
\begin{equation}
\lim_{\nu \to \infty}
Q_{ \omega_L = \nu }\sim {\Gamma \Omega^2 \over  2 R \left( \Gamma^2 + 2 \Omega^2 \right) }
\label{eqnu1}
\end{equation}
a super-Poissonian behavior.
In the {\em intermediate modulation limit},
when $R = \Gamma$, we obtain
\begin{equation}
 \lim_{\nu \to \infty}
 Q_{\omega_L = \nu} = - { 81 \Omega^2 \Gamma^2 \over  2  \left( 9\Gamma^2
 + 4 \Omega^2 \right)^2 }
\label{eqnu2}
\end{equation}
a sub-Poissonian behavior.
When $R \to \infty$ we find that $Q$ is small 
\begin{equation}
 \lim_{\nu \to \infty}
 Q_{\omega_L = \nu} \sim - { \Omega^2  \over  2 R \Gamma }.
\label{eqnu3}
\end{equation}
Eqs. (\ref{eqnu1}-\ref{eqnu3}) are valid only
in the limit of $\nu \to \infty$. However behaviors
similar to the predictions of these equations
are found also for finite values
of $\nu$. On $\omega_L=\nu$ we have three typical
behaviors: $(i)$ $Q(\omega_L = \nu) > 0$ when the process is slow, see
Fig. \ref{fig1}). $(ii)$ $Q(\omega_L = \nu)<0$ when $\Gamma \simeq R$, see Fig.
\ref{fig4} for $R= \Gamma/2$, and $(iii)$ when $R \to \infty$ we find
$Q \left( \omega_L = \nu \right) \to 0$, Fig. \ref{fig3}. 

\begin{figure}
\begin{center}
\epsfxsize=70mm
\epsfbox{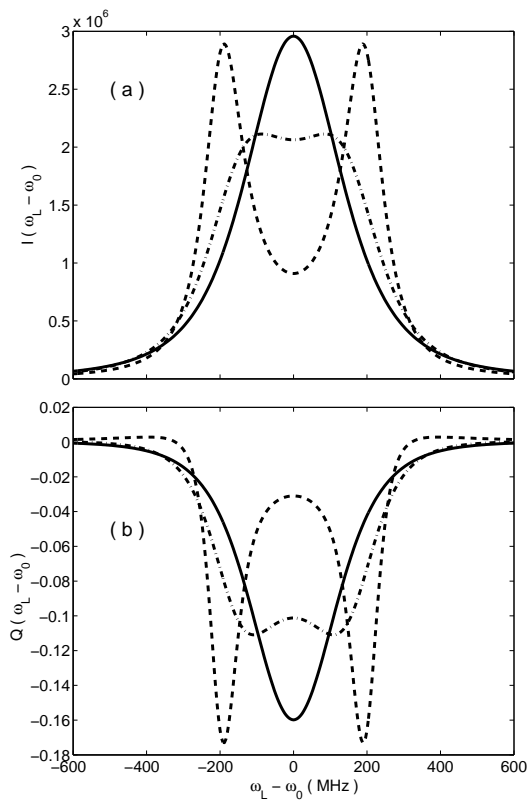}
\end{center}
\caption
{Sub - Poissonian behaviors of $Q$ for the case
of strong coupling $\nu>>R$ and intermediate modulation $R\simeq
\Gamma \simeq \Omega$ limit.
$Q$ exhibits two types of quantum sub-Poissonian statistics:
i) splitting behavior and ii) motional narrowing behavior
where $Q$ has a single minimum.
The parameters are:
$\Gamma = 40$ MHz, $\nu = 5 \Gamma $, $ \Omega = \Gamma / \sqrt{2} $,
and $ R = 1 \Gamma $ (dashed curve), $ 3 \Gamma $ (dashed-dot curve), and $ 5 \Gamma $ (solid curve).
}
\label{fig5}
\end{figure}

\subsection{Intermediate Modulation Limit $R \simeq \Gamma$ 
}

 When $R \simeq \Gamma \simeq \Omega<<\nu$
we obtain interesting behaviors
for $Q$.  In Fig. \ref{fig4} the $Q$ parameter
shows a transition from sub-Poissonian to super Poissonian,
photon statistics.
In this regime of parameters, the shape of $Q$ when plotted
as  a function of $\omega_L$ is very sensitive
to the value
of the control parameters e.g.  in Fig. \ref{fig4} we change
 $R$ only moderately still we see very different
types of behaviors for $Q$.
 For certain values of parameters
$Q$ attains more than two peaks (see Fig. \ref{fig4} for
$R=\Gamma/6,\Gamma/4$).
In contrast $I(\omega_L)$ exhibits a simple
splitting behavior with two peaks on $\pm \nu$,
which is similar to the slow modulation case.

 Besides the transition from sub to super Poissonian
 behavior, a second type of transition is observed
 as $R$ is increased. In our problem we have
 two types of sub-Poissonian behavior.
 We noticed
 already that when  the stochastic modulation becomes
 very fast, $Q$ has one minimum on zero detuning
 (see Fig.  \ref{fig2}), while when $R \simeq \Gamma$,
 $Q$ has two minima on $\omega_L \pm \nu$ 
 (see Fig. \ref{fig3} and $R=\Gamma/2)$.
 The transition between these two types of sub-Poissonian
 behaviors is shown in Fig. \ref{fig5}.   

\begin{figure}
%\begin{center}
%\includegraphics[totalheight=6.5in]{figb17.eps}
%\includegraphics[totalheight=6.5in]{fig6.eps}
%\end{center}
\begin{center}
\epsfxsize=70mm
\epsfbox{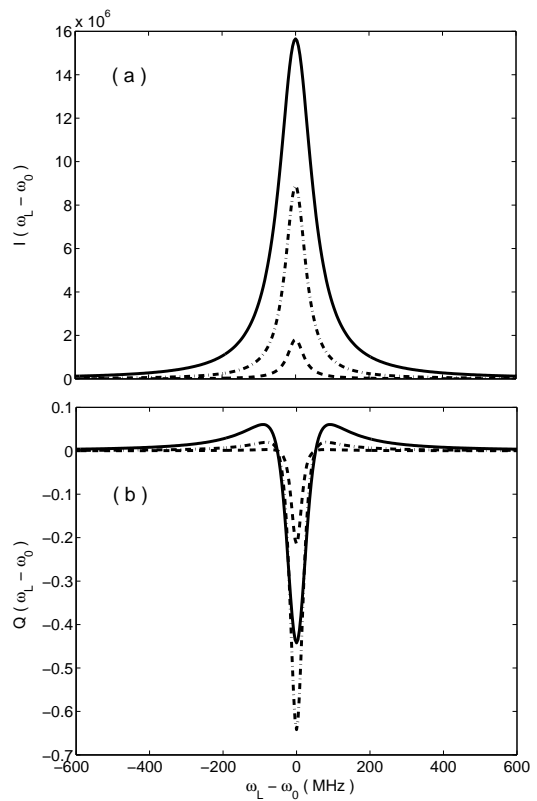}
\end{center}
\caption
{We investigate the dependence the line shape and $Q$
on the Rabi frequency $\Omega$.
Power broadening of the line is observed, while
for 
the $Q$ parameter we have a turn-over behavior
as explained in the text.
The three values of the  Rabi frequency are 
$\Omega = \Gamma /4  $ (dashed curve), 
$ \Gamma  / \sqrt{2}$ (dashed-dot curve), and $ 3 \Gamma / 2  $
(solid curve).
And the minimum of $Q$, on zero detuning,  is obtained for 
the intermediate value of $\Omega =\Gamma/\sqrt{2} $.
The fixed parameters 
are $\Gamma = 40$ MHz, $\nu = 5 \Gamma $, 
and $R =100 \Gamma$.
}
\label{fig6}
\end{figure}
%\subsection{Dependence of Excitation Field}
%  We know that the line shape and the $Q$ parameter are the function of Rabi frequency.
%Rabi frequency is directly proportional to the strength of excitation field.  
%as the Rabi frequency increases, the line shape is extended, and the $Q$ parameter 
%also changes obviously. In Fig.\ref{figb17}, we take three values of the Rabi frequency,
%$ \Omega = \Gamma /4  $ (dashed curve),$ \Gamma  / \sqrt(2)$ (dashed-dot curve), 
%and $ 3 \Gamma / 2  $(solid curve), and the $R$ fixed at$ 100 \Gamma$ in the fast modulation regime.
%As the Rabi frequency increases, the line shape is monotonously extended. The $Q$ parameter is
%small at first, then gets to minimum, and finally return small value at$ \omega_L = 0 $.
%  The sub-Poissonian statistics $( Q < 0 )$ is very interested because it is an explicit
%feature of quantum field. One can chooses the Rabi frequency in order to obtain strong 
%sub-Poissonian in experiment. 
%For zero detuning, $ \omega_L = 0 $, we obtain from Eq. (\ref{eqQ})
%
%\begin{equation}
%Q =
%-{2\,{\Gamma \Omega}^2\,\left[ 4\,{\nu}^2\,\left( -\Gamma + 4\,R \right)  +
%      3 \Gamma,\left( \Gamma + 4\,R \right)^2 \right] \over \left[ 4\,\Gamma {\nu}^2 +
%       \left( {\Gamma}^2 + 2\Omega^2 \right) \,\left( \Gamma + 4\,R \right)
%\right]^2}.
%\label{refzdet}
%\end{equation}
%
\begin{figure}
%\begin{center}
%\includegraphics[totalheight=6.5in]{figb13.eps}
%\includegraphics[totalheight=6.5in]{fig7.eps}
%\end{center}
\begin{center}
\epsfxsize=70mm
\epsfbox{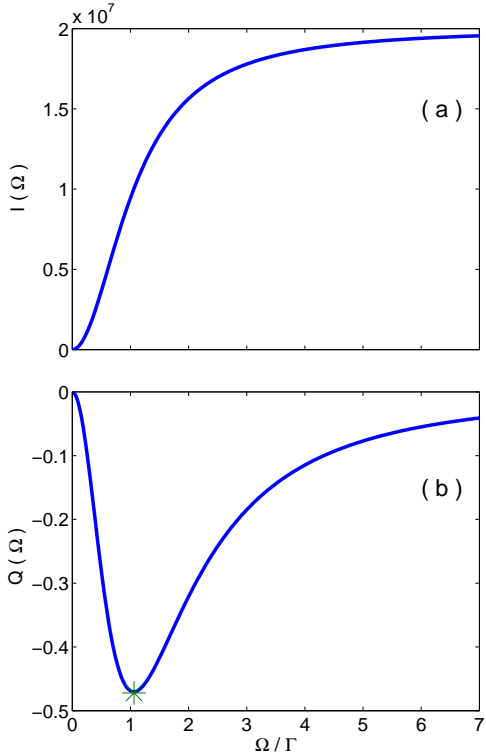}
\end{center}
\caption
{
The photon emission rate  $I(\Omega)$ (a) the $Q(\Omega)$ parameter (b)
are calculated with the exact solution of Eq. (\ref{eqLS})
 and Eq. (\ref{eqQ}),
in the fast modulation limit. 
The parameters are  
$\Gamma = 40$ MHz, $\nu = 5 \Gamma $, $\omega_L=0$, $ \Omega = 0 \to 7 \Gamma $,
and $ R = 10 \Gamma $.
We observe the saturation of
the line as $\Omega$ 
becomes large, while $Q(\Omega)$ has a minimum. 
The value of $\Omega$ which minimizes $Q(\Omega)$ is
of interest since it yields the strongest quantum
fluctuations. 
The symbol $*$ is the point  $(\Omega_{\mbox{min}},Q_{\mbox{min}})$
calculated based on the approximation Eqs. (\ref{eqsim1},\ref{eqsim2}).
%\Omega_{min} = 1.1 \Gamma, and Q_{min} = - 0.47.
}
\label{fig8}
\end{figure}

\begin{figure}
\begin{center}
\epsfxsize=70mm
\epsfbox{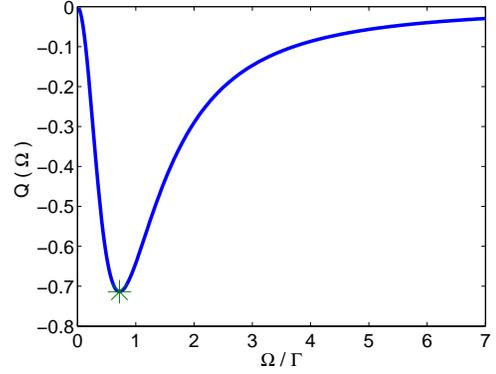}
\end{center}
\caption
{
We demonstrate how to choose the Rabi 
frequency in order to obtain strong sub-Poissonian behavior.
The $Q(\Omega)$ parameter 
is calculated with the exact solution of Eq.
(\ref{eqQ}).
The detuning  $\omega_L - \omega_0$ is zero, 
and  $\Gamma = 40$ MHz, $\nu = \Gamma / 10 $
and $ R = \Gamma / 100$  corresponding to
the slow modulation weak coupling limit.
The minimum, the $*$ in the figure,
$Q_{\mbox{min}}\simeq -0.71$ is found for  $\Omega_{\mbox{min}} \simeq
 0.72 \Gamma$, which is close to the global minimum
$(Q_{\mbox{min}},\Omega_{\mbox{min}})=(-3/4,\Gamma/\sqrt{2})$.
}
\label{fig7}
\end{figure}

\section{Extremum of  $Q$}

 We now investigate the dependence of $Q$ on the excitation
field. In Fig. \ref{fig6} we consider an example
line shape and $Q$ parameter, where we fix
the model parameters $\nu,R, \Gamma$ and vary the Rabi
frequency. 
For the line we see well known power broadening:
as the Rabi frequency is increased the line becomes
wider, and as expected the photon emission rate $I(\omega_L)$
increases monotonically when $\Omega$ is increased.
For the $Q$ parameter we have a turn-over behavior,
as we increase $\Omega$ the value of
$Q$ on zero detuning decreases then
increases.

 Generally this type of turn-over is expected, since as discussed
in Sec. \ref{SecIntroSub},
$Q =0$ when
$\Omega \to \infty$ or $\Omega \to 0$. Thus there exists 
an optimal Rabi frequency which yields an extremum of $Q$. 
 Obviously it is important to obtain the values of $\Omega$ which
yield the extremum of $Q$, since then the fluctuations  are
the largest. The extremum can be either 
a minimum or a maximum, as we shall show now. 

\subsection{Largest Quantum Fluctuations}

 We now consider the quantum regime $Q<0$.
In Fig. \ref{fig8}(b) we demonstrate the turn-over behavior of $Q(\Omega)$
 for
an example where the stochastic fluctuations are fast.
In this fast modulation case $Q<0$ hence $Q(\Omega)$ has a minimum.
For the same parameters 
the photon emission rate $I(\Omega)$
saturates as $\Omega$ is increased, and the 
emission rate is never faster than $\Gamma$ [see Fig. \ref{fig8}(a)].

Let $\Omega_{\mbox{min}}$ be the Rabi frequency which
minimizes $Q$ in the sub-Poissonian case $Q<0$, and
$Q_{\mbox{min}}$ the corresponding value of $Q$.
Using Eq. 
(\ref{eqZT})
we find for zero detuning,
and for $R > \Gamma/4$
\begin{equation}
\Omega_{\mbox{min}} = \sqrt{  \Gamma^3 + 4 \Gamma^2 R + 4 \Gamma \nu^2  \over 2 \left( \Gamma + 4 R \right) }
\label{eqmin1}
\end{equation}
and
\begin{equation}
Q_{\mbox{min}} = 
\frac{-\left( 4\,{\nu}^2\,\left( -\Gamma + 4\,R \right)  +
      3\,\Gamma\,{\left( \Gamma + 4\,R \right) }^2 \right) }{4\,
\left( \Gamma + 4\,R \right) \,\left( \Gamma^2 + 4\,{\nu}^2 + 4\,\Gamma\,R \right) } .
\label{eqmin2}
\end{equation}

 Eqs. (\ref{eqmin1},\ref{eqmin2})
yield $\Omega_{\mbox{min}}$ and $Q_{\mbox{min}}$
in terms of $\nu$ and $R$.
 Due to motional narrowing effect, 
for fast processes satisfying $\nu << R$,
$\nu$ and $R$ are not easily obtained from experiment,
while the parameter $\Gamma_{{\rm SD}}$
is in principle easy to  obtain from the measurement of
the line width. 
Using Eq.
(\ref{eqQ3}), we find in the fast modulation limit
and for zero detuning 
\begin{equation}
{\Omega_{\mbox{min}} \over \Gamma} = 
\sqrt{{ 1  + \Gamma_{\rm{SD}}/\Gamma \over 2} }
\label{eqsim1}
\end{equation}
\begin{equation}
Q_{\mbox{min}} = -  {\left( \Gamma_{\rm{SD} } + 3 \Gamma \right) \over
4 \left( \Gamma + \Gamma_{SD} \right)
}
\label{eqsim2}
\end{equation}
These simple equations  relate between the
width of the line given in Eq. 
(\ref{eqLS3a})
and $Q_{\mbox{min}}$  
and $\Omega_{\mbox{min}}$.  
From Eqs. (\ref{eqsim1},\ref{eqsim2}) we see that when $\Gamma_{SD}<<\Gamma$,
$\Omega_{\mbox{min}}=\Gamma/\sqrt{2}$ and $Q_{\mbox{min}}=-3/4$.
When $\Gamma_{SD}>>\Gamma$ we find 
 $\Omega_{\mbox{min}}=\sqrt{\Gamma_{SD}\Gamma/2}$ and
$Q_{\mbox{min}}=-1/4$.

{\bf Remark:}
When the detuning is not zero,
we find using Eq. (\ref{eqQ3})
\begin{equation} 
{\Omega_{\mbox{min}} \over \Gamma} = \sqrt{
{\left( 1 + \Gamma_{{\rm SD}}/\Gamma \right)^2 + 4 \omega_L ^2/\Gamma^2 \over
2 \left( 1 + \Gamma_{{\rm SD}}/\Gamma \right) }.
}
\end{equation}

\begin{figure}
\begin{center}
\epsfxsize=70mm
\epsfbox{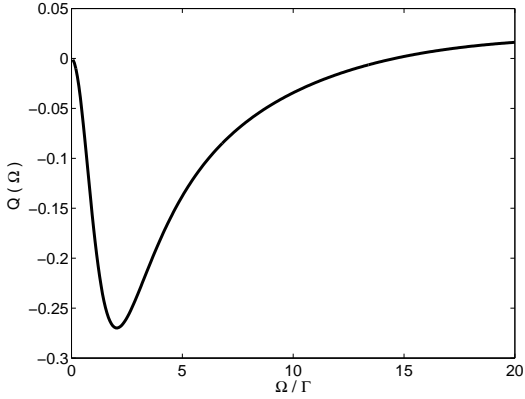}
\end{center}
\caption
{
Same as Fig. \ref{fig7} for the intermediate modulation
limit $\Gamma \simeq R$. Unlike Figs. \ref{fig8},\ref{fig7},
the detuning is  on $\omega_L =\nu$. 
Parameters are chosen as $\Gamma = 40$ MHz, 
$\nu = 5 \Gamma $, $ \Omega = 0 \to 20 \Gamma $,
and $ R = 2 \Gamma $.
The figure illustrates that the turn-over behavior of
$Q(\Omega)$ is generic, and not limited to 
the  fast modulation limits Fig. \protect{\ref{fig8}} 
and slow modulation/weak coupling limit Fig. \ref{fig7}.
}
\label{fig9}
\end{figure}

 Similar turn-over behaviors of $Q(\Omega)$ are found also in 
other non fast parameter regimes.
In Fig. \ref{fig7}
we show $Q(\Omega)$ 
versus
$\Omega$ for the slow modulation weak coupling limit
$R <\nu < \Gamma$ and for zero detuning.
Fig. \ref{fig7} shows that $Q$ exhibits a minimum as function of $\Omega$,
this minimum is found in the vicinity of 
$\Omega_{\mbox{min}}= \Gamma/\sqrt{2}$.
Such a behavior is understood based on
Eq. (\ref{eqQapp}), the spectral diffusion contribution
for $Q$ is not important at zero detuning, while
the contribution of $Q_{M}$ yields $\Omega_{\mbox{min}}=\Gamma/\sqrt{2}$.
To demonstrate that the turnover behavior of 
$Q(\Omega)$ is generic, we consider also
the intermediate modulation limit in Fig. \ref{fig9}.
Here we choose the detuning according to $\omega_L = \nu$,
since the $Q$ parameter on zero detuning is relatively
small 
(see Fig. \ref{fig5}).

\subsection{Maximum of Super-Poissonian Fluctuations}

\begin{figure}
\begin{center}
\epsfxsize=70mm
\epsfbox{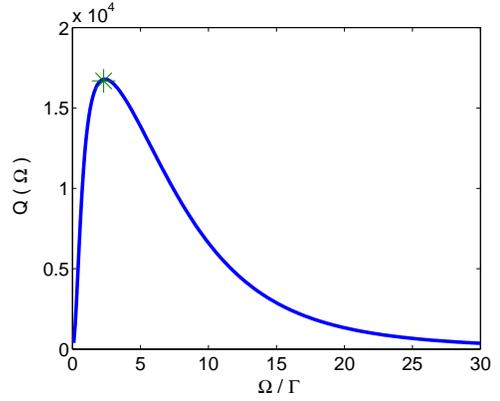}
\end{center}
\caption
{
The crossover behavior of the $Q\left( \Omega \right)$ parameter in the slow
modulation limit. $Q$ exhibits a super-Poissonian behavior,
and $Q\left( \Omega \right)$
attains a maximum when the Rabi frequency is changed.
Thus also for this classical type of behavior an extremum
of the fluctuations is found for a particular
value of the control parameter $\Omega$.
The parameters are
$\Gamma=40$ MHz,
$\nu=5 \Gamma$, $R=500$Hz and
$\omega_L-\omega_0 = \nu$.
Using the approximate  Eqs. (\ref{eqyh2},\ref{eqyh3}) 
we obtain the values $Q_{\mbox{max}}\simeq 16686$
$\Omega_{\mbox{max}} \simeq  2.3 \Gamma$, 
the symbol $*$ in the figure, which is in
good agreement with the exact solution (the solid curve).   
}
\label{fig10}
\end{figure}

In contrast to the behaviors in the quantum regime $Q<0$,
in the slow modulation limit where $Q>0$, $Q(\Omega)$ 
obtains a maximum, whose location is easy to  calculate
with Eq.  
(\ref{eqQ1}). 
 Such a behavior is demonstrated in  Fig. \ref{fig10}
for a case when the spectral shift $\nu$ is not very large.
If $\nu >> \Omega, \Gamma$ then in the slow modulation limit
\begin{equation}
Q_{\mbox{slow}} \simeq 
{ \Gamma \Omega^2 \over 2 R \left(\Gamma^2 + 2 \Omega^2 \right) }
               -   {  3 \Gamma \Omega^2 \over 32 R \nu^2  }
\label{eqyh1}
\end{equation}
when the detuning is equal to $\omega_L = \nu$.
In Eq. (\ref{eqyh1}) the second term on the right hand side
is supposed to be a correction to the first term, namely $Q_{\mbox{slow}}>0$.
Let $\Omega_{\mbox{max}}$ be the value of $\Omega$ which
maximizes  $Q$ in the super-Poissonian regime,
and $Q_{\mbox{max}}$ the corresponding maximum.
This  maximum always exists since as mentioned $Q=0$ when
$\Omega \to 0$ or $\Omega \to \infty$. 
Then using Eq. (\ref{eqyh1})
\begin{equation}
\Omega_{\mbox{max}} \simeq 
\sqrt{{ \Gamma \left(4 \sqrt{3} \nu - 3 \Gamma \right) \over 6
}}
\label{eqyh2}
\end{equation}
which is independent of $R$
and
\begin{equation}
Q_{\mbox{max}} \simeq { \Gamma \left( 3 \Gamma^2 + 16 \nu^2 - 8 \sqrt{3} \Gamma \nu \right)
            \over 64 R \nu^2}.
\label{eqyh3}
\end{equation}
Note that when the frequency shifts are 
very large $\nu\to \infty$ we find
using Eq. (\ref{eqyh2})
$\Omega_{\mbox{max}} \to \infty$. 
Hence the value of 
$\Omega_{\mbox{max}}$ may become very large and then
in experiment it is impossible to reach $\Omega_{\mbox{max}}$
(e.g. $\nu=1GHz$). 
If we impose the condition $\Omega<<\nu$ 
we have 
\begin{equation}
Q_{\mbox{slow}} 
\simeq { \Gamma \Omega^2 \over 2 R \left( \Gamma^2 + 2 \Omega^2\right) }
\end{equation}
for the laser detuning  $\omega_L=\nu$. 
Hence $Q_{\mbox{slow}}$ monotonically increases
and eventually  saturates, similar to the behavior
of the average emission rate. 

\section{Summary and Discussion}

The $Q$ parameter yields informations not contained
in the line shape.
The most obvious is the 
transition from super (i.e. classical) to sub (quantum)
Poissonian behavior. Such a quantum
signature of the photon emission process is not obtained from
the line shape. In comparison with the $Q$ parameter of
a single atomic transition, the $Q$ parameter investigated
here
exhibits rich behaviors. These include splitting, both
in the sub and in the super Poissonian regime, a transition
from a fast to a slow modulation limit, and
effects related to motional narrowing. 
The most non-trivial behavior is obtained
in the intermediate modulation  limit when $\Gamma\simeq R \simeq \Omega$
where $Q$ attains more then two peaks. 
 
 Since $Q$  contains the new information on single
molecule experiments, namely information beyond the
line shape, it is important to
emphasize that $Q$ attains an extremum for a
particular value of  the Rabi
frequency. In particular  in the sub-Poissonian
regime $Q(\Omega)$ has a minimum.
Hence we optimize the Rabi frequency
in such a way that $|Q|$ is increased,
e.g. we obtain $\Omega_{\mbox{min}}$.   
In other words there exist an ``ideal'' choice of the Rabi frequency
in single molecule experiments. 
In the quantum sub-Poissonian regime
this optimal Rabi frequency cannot be considered weak,
neither strong, hence perturbative  approaches to single molecule spectroscopy
are not likely to yield it.
This is in complete contrast to most theories of line
shapes which are based on the assumption of weak external
fields, e.g. the Wiener--Khintchine theorem
and linear response theory.
Single molecule theories should be able to predict the turn-over
behavior of $Q(\Omega)$ based on
different models, since such a behavior is not expected
to be limited to the model under  investigation.
Of-course the exact solution presented in this manuscript
is very valuable in this direction,
since it predicts precisely the details of this transition
for the Kubo-Anderson stochastic process. 

 It would be interesting to investigate further 
how general are our results. From line shape theory,
we know that in the fast modulation limit, line
shapes have  Lorentzian shapes under very general
conditions. 
From experiment we know that motional narrowing effect,
and Lorentzian behavior of lines 
is wide spread. 
Thus at-least in this limit certain general features
of line shapes, which are not sensitive to model
assumptions are found. 
Similarly, we expect, that in the fast modulation
 limit, some of our results are general. For example
the motional narrowing behavior of $Q$ and its approach
to Mandel's behavior $Q_M$ is likely to be general.
It
 would be interesting to check if  the relation between
$Q_{\mbox{min}}$ and $\Omega_{\mbox{min}}$ 
Eq. (\ref{eqsim1},\ref{eqsim2}), and the width of the line
given by $\Gamma_{SD}$ and $\Gamma$  is valid for other
models, both stochastic and Hamiltonian. These simple
equations are important since
they yield the optimal Rabi frequency
$\Omega_{\mbox{min}}$ in terms of  the width
of the line shape, which in turn is easily determined  in usual
line shape measurement.  

 In Ref. (\cite{PRL},\cite{ADV}) a semi-classical 
framework for the mathematical calculation
of $Q$ for single molecule spectroscopy was investigated.
The approach yields the $Q$ parameter in terms of a Fourier
transform of a three
time dipole correlation function. 
As mentioned in the introduction,
the approach in  (\cite{PRL},\cite{ADV}) is based on
two main approximations (i) external fields
are weak $\Omega \to 0$, i.e. linear response theory, 
and  the (ii) semi-classical approach to photon
counting statistics. The second assumption implies that
$Q>0$, and as pointed out in \cite{BarkaiRev,ADV} such an approach is
expected to be valid for slow processes.
The approach is useful since most single molecule experiments
report on slow fluctuations. 
The results obtained in this manuscript reduce
to those in (\cite{PRL},\cite{ADV}) in the limit of 
$\Omega \to 0$, and in the slow modulation limit,
as they should.  
 The quantum behavior of $Q$ becomes
important when $R \sim \Gamma$ or for faster
processes.  
It is left for future
work to construct a general
quantum linear
response theory, based on the Eqs. of motion 
\ref{eqmtz},
which would yield
both super and sub-Poissonian statistics. 
Finally, also the investigation of the time dependence of
$Q$ is timely. 

This work was supported by National Science Foundation award CHE-0344930.
EB thanks the Complexity Center in Jerusalem for financial
support.

\end{document}